# A Comparison between Separately Calibrated $P$-$\alpha$ and Mesoscale Models for Weak Shock Compaction of Granular Sugar

Dawa Seo,[1,2,a] Darby J. Luscher,[3,b] and Nitin Daphalapurkar[2,c]

[1]*Department of Civil and Environmental Engineering, Wayne State University*

[2]*Theoretical Division, Los Alamos National Laboratory*

[3]*X Computational Physics Division, Los Alamos National Laboratory*

This study compares calibration strategies for predicting particle velocity in granular sugar subjected to weak shock loading, using measurements from flyer-plate impact experiments as a benchmark. Two computational approaches are evaluated: a continuum-based $P - \alpha$ Menikoff model requiring calibration of effective constitutive parameters, and mesoscale simulations that explicitly resolve grain geometry and porosity. Both models can match the measured particle-velocity histories, but only through fundamentally different calibration mechanisms. In the $P - \alpha$ model, a pressure-dependent yield strength is essential and the response remains highly sensitive to choices of parameters such as the crush-out pressure. In contrast, mesoscale simulations are far less sensitive to parameter tuning and instead depend primarily on the physical state variable of porosity, represented in 2D through an equivalent mapping of the 3D specimen. These results show that continuum parameters act as effective surrogates for underlying grain-scale processes, whereas mesoscale modeling identifies porosity as the dominant control on macroscopic wave onset, highlighting distinct calibration pathways and interpretive implications for each modeling approach.

[a]Electronic mail: dawaseo@wayne.edu

[b]Electronic mail: djl@lanl.gov

[c]Electronic mail: nitin@lanl.gov

Core principles and modeling approaches for shock-compacted granular materials have often simplified systems to single-component representations with averaged grain properties like porosity. Continuum, equation-of-state-based porous-media models are widely used to simulate the crushing response of granular materials under shock loading. For example, the $P$ -$\alpha$ model describes the Hugoniot locus between shock and particle velocities for ductile porous materials, as shown in Eq 1[1].

$$P = f(V_S, E) = f\left(\frac{V}{\alpha}, E\right), \quad \text{where } \alpha = \frac{V}{V_S}. \tag{1}$$

This formulation characterizes the equation of state using specific volume and internal energy, $E$, with distension $\alpha = V/V_s = \rho_s/\rho$, where $V$ and $\rho$ are the porous (bulk) specific volume and density and $V_s$ and $\rho_s$ are the corresponding matrix values. This captures effective pore-scale responses, such as deformation, displacement, and rotation, which can significantly affect the bulk behavior[2].

In addition, there is a significant uncertainty in the analytic form of the shock versus particle velocities ($U_s$-$U_p$) for granular materials, especially in the weak shock regime[3]. This is because both hydrodynamics and strength effects can be important for accurately modeling the propagation of the weak shock[4]. To accurately describe the mechanical response of the heterogeneous granular systems, mesoscale models have been introduced, an intermediate length scale between macro and micro. Mesoscale modeling enables investigation of particle-scale responses that continuum-based models cannot capture.

This paper compares calibration approaches for particle-velocity predictions from a continuum-based $P$ -$\alpha$ Menko model and a mesoscale model for shock-loaded granular sugar, focusing on a single weak-shock condition (295 m/s) to enable controlled comparison of the mechanisms represented in each approach. A parametric study identifies the best fit to gas-gun flyer-plate measurements, and analysis of particle-velocity, porosity, and pressure evolution provides guidance for effective use of the $P$ -$\alpha$ model with mesoscale results serving as the baseline.

This study utilizes the gas gun experimental setup of Sheffield, Gustavsen, and Alcon. (1998)[5], illustrated schematically in Fig. 1a with additional details provided in the Supplementary. The compacted sugar layer had a density of 65% of its theoretical maximum (TMD). The tests were carried out at an impact velocity of 295 $m/s$. They used magnetic-velocity gauges to measure the particle velocities traveling from the impactor to backing plug through granular sugar shown in Fig. 1a, determining the shock input and wave dispersion response.

Both P-$\alpha$ and mesoscale models were developed and simulated in FLAG hydrocode, a multiphysics hydrocode developed at Los Alamos National Laboratory. This hydrocode, based on the Lagrangian finite volume method solves the deformations in space at each simulation time (details in the Supplementary). For both models, Mie-Gruneisen Equation of State was employed to calculate the thermodynamic response between pressure, density, temperature, and internal energy. The continuum model uses a homogenized domain with porosity as an internal variable to capture the compaction, while the mesoscale model explicitly simulates particle interactions and pore crushing of a distribution of pore spaces numerically constructed based on dry pluviation or consolidation technique. In the simulation, particle velocity was recorded at the front and back face centers, aligned with the experimental setup's line of sight (details in the Supplementary).

Granular sugar (sucrose) has attracted interest as an inert surrogate of the Cyclotetramethylenetetranitramine (HMX) due to its similar particle size distribution to HMX, an organic material, and its availability. In this study, we reference the experiments conducted by Sheffield, Gustavsen, and Alcon. (1998)[5] to build and calibrate both continuum-based and mesoscale models with granular sugar, as shown in Fig. 1a. Material properties for both computational modeling were selected from the previous studies, as listed in Table I.

TABLE I. Material properties

| Property | Sugar[6,7,5] | Kel-F[8] | PMMA[9] |
|---|---|---|---|
| Density, $\rho$ [$g/cm^3$] | 1.5805 | 2.02 | 1.182 |
| Gruneisen Parameter, $\Gamma_0$ | 1.04 | 0.4 | 0.85 |
| Sound Speed, $c_0$ [$km/s$] | 3.04 | 1.772 | 0.218 |
| Yield Stress, $\sigma_y$ [$kbar$] | 1.1 | 0.167 | 0.42 |
| Shear Modulus, $sm_0$ [$Mbar$] | 0.1 | 0.0126 | 0.0232 |

For the material domain, P-$\alpha$ model establishes a continuum domain for the granular system to evaluate materials response. Mesoscale model generates the domain of individual grains, and their boundary controls the interaction with other grains to transmit forces and prevent interpenetration between deformable grains[4].

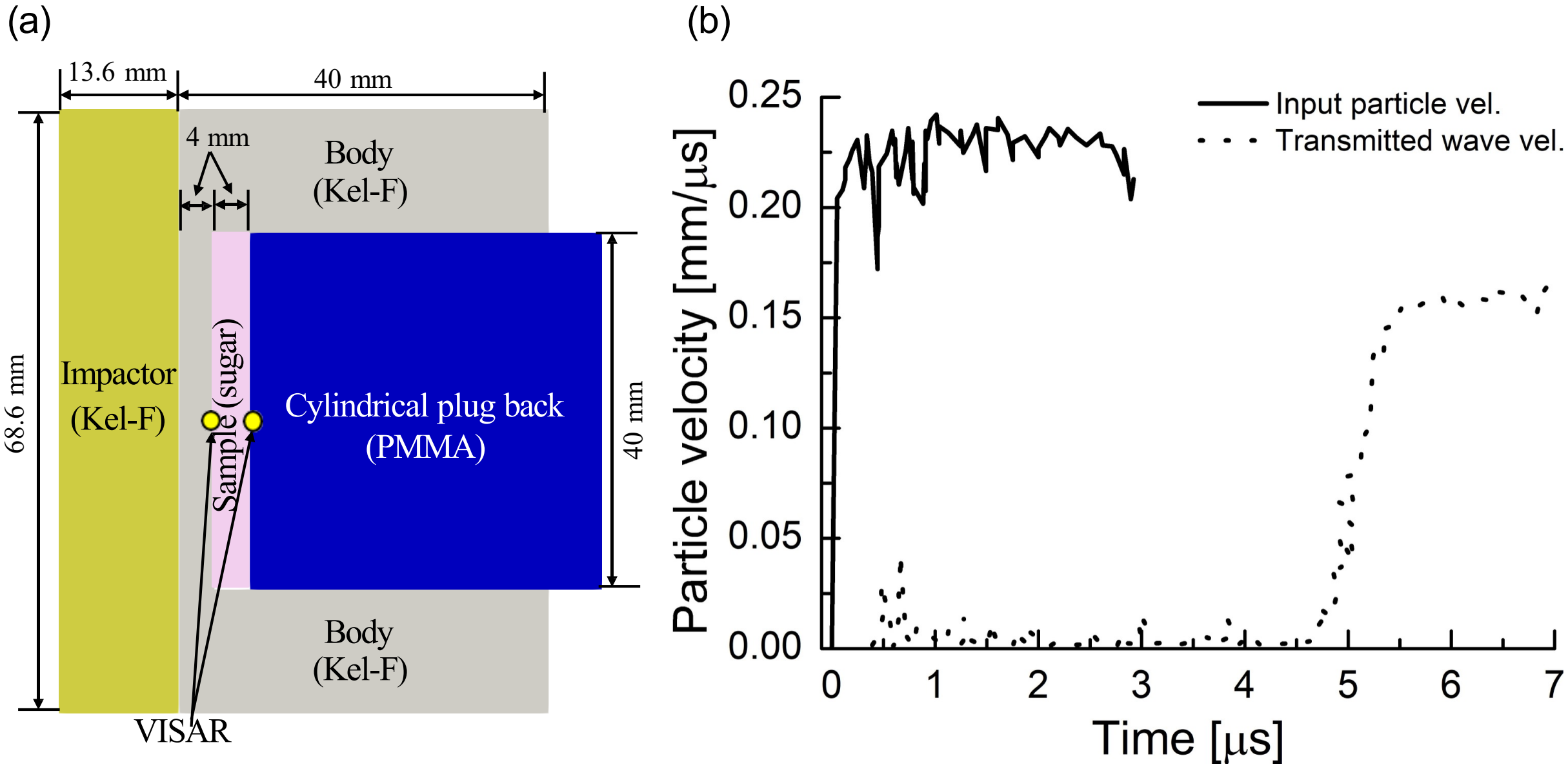

FIG. 1. (a) Setup for computational modeling (b) Experimental data, reproduced from S. A. Sheffield, R. L. Gustavsen, and R. R. Alcon, AIP Conf. Proc. 429, 575–578 (1998)[5], with the permission of AIP Publishing

This study applied the $P$ -$\alpha$ Menko model suggested by Menikoff and Kober (2000)[10]. They reformulated the $P$ -$\alpha$ model within a thermodynamically consistent framework, making it applicable to other materials that exhibit concave Hugoniot loci in (particle velocity, shock velocity)-plane. Eq. 2 shows the Menikoff and Kober model (MK),

$$P_s v_s = P_{s,\mathrm{ref}}\, v_{s,\mathrm{ref}} - P_c\, v_{s,\mathrm{ref}} \log\left(\frac{1-\chi_{\mathrm{eq}}}{1-\chi_{\mathrm{ref}}}\right) \quad (2)$$

where solid pressure $P_s$, solid specific volume $v_s$, and solid volume fraction $\chi$. Menikoff and Kober[10] use $\phi$ for the solid volume fraction; our $\chi$ corresponds to their $\phi$. Accordingly, our porosity (void fraction) is $1-\chi$ and the distension is $\alpha = 1/\chi$. Variables with 'ref' denote a reference state, while those with 'eq' refer to equilibrium conditions. In this work, the reference-state quantities ($P_{s,ref}$ ,$v_{s,ref}$ , $\chi_{ref}$ ) are taken from the initial porous configuration from which compaction begins. In this work we denote Menikoff-Kober's compaction pressure scale by $P_c$ (instead of $P_{comp}$) to align with the "crush-out pressure" terminology used in our FLAG input. Here $P_c$ is a characteristic compaction pressure scale that controls the rate and curvature of porosity reduction in the smooth Menikoff-Kober compaction law. Although termed a *crush-out pressure*, $P_c$ is not a sharp cutoff for complete pore collapse;

rather, it governs the gradual approach to the fully dense limit.

This study used dry pluviation to generate contacting square grains (0.120 mm equivalent diameter) as shown in Fig. 2a. Although 2D simulations cannot fully capture the 3D contact network, they have been widely used for the lower computational cost and ability to reproduce key trends[11]. Accordingly, we use an equivalent-porosity 2D packing that preserves volumetric compaction behavior and has been validated for shock compaction[4] (more details in Supplementary). Porosity was statistically represented to account for 2D simulation of 3D experiments using Eq. 3, where $D$ is relative density and $n$, $n_{max}$, and $n_{min}$ are porosity values.

$$D = \frac{(n_{max} - n)}{(n_{max} - n_{min})} \tag{3}$$

Theoretical porosity ranges (minimum and maximum) are 0-0.5 in 2D, and 0-0.66 in 3D. A target 2D porosity of 0.27 was chosen to match the experimental 3D porosity of 0.35 at equal relative densities. Mesoscale properties matched those of $P$ -$\alpha$ (Table I), assuming plastic deformation captures the shock response for brittle granular sugar without fracture and friction (further details in Supplementary). Under dynamic shock loading, the dominant macroscopic response arises from pore collapse and grain compaction, which suppress large-scale crack propagation. As such, the brittle grains were approximated by an elasto-plastic response, consistent with previous mesoscale studies of crushed brittle solids under confinement, and proven to be satisfactory[4,12]. This assumption allows the dominant inelastic behavior to be captured while maintaining computational tractability.

Figures 2b-e presents representative simulation results depicting pressure distributions under an impact velocity of 295 m/s, generated using both the $P$ -$\alpha$ model and the mesoscale model. The zone pressure, $z_p$, refers to the pressure associated with a finite volume within the material domain. In the $P$ -$\alpha$ model, pressure propagates uniformly through the granular sugar to its rear boundary (Fig. 2d). In contrast, the mesoscale model reveals a heterogeneous shock compaction front extending from the granular sugar to the PMMA interface, as shown in Fig. 2e. This contrast underscores a fundamental distinction between continuum and granular materials, namely, that mesostructural variability in granular media plays a critical role in shaping local stress distribution.

To obtain the best fit of the $P$–$\alpha$ Menko model to the experimental particle velocity data from Sheffield, Gustavsen, and Alcon. (1998)[5], we used literature-based material properties

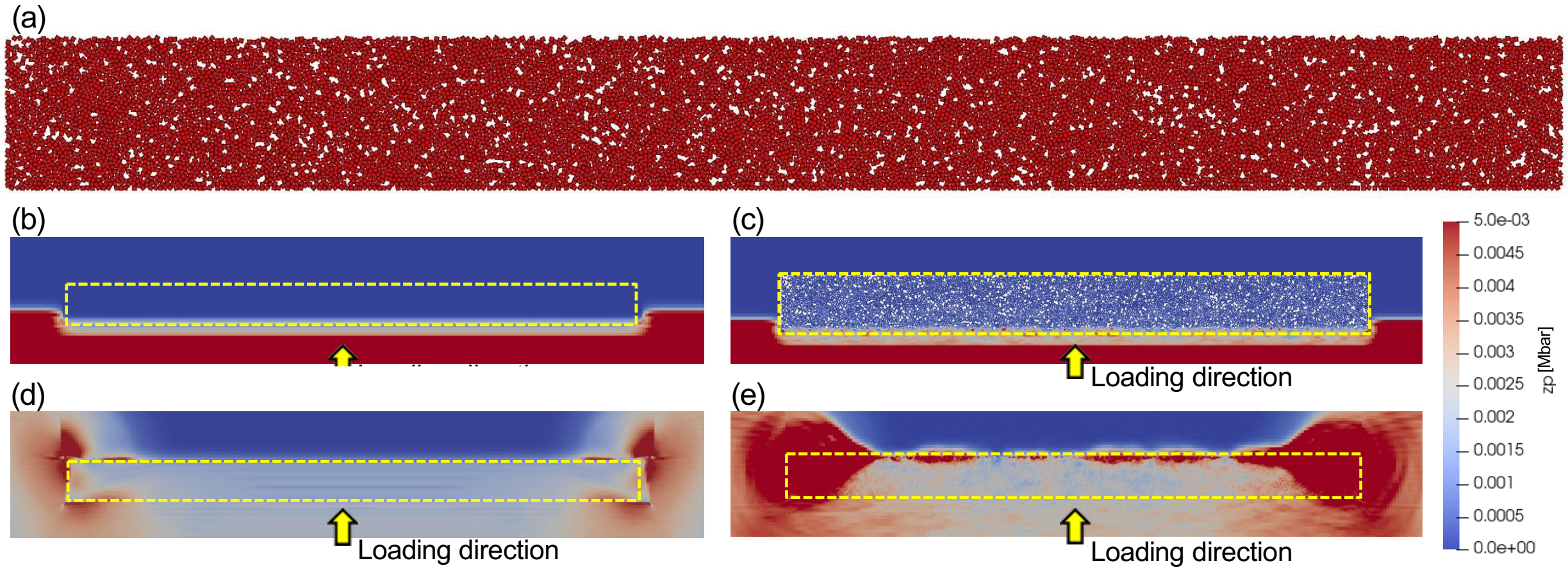

FIG. 2. (a) Testing granular materials prepared by dry pluviation. Zone pressure results from $P$-$\alpha$ model and mesoscale simulations, respectively. The onset of the plateau in input particle velocity occurs at (b) 2.72 $\mu$sec and (c) 2.4 $\mu$sec, while the transmitted velocity reaches its plateau at (d) 7.12 $\mu$sec and (e) 6.83 $\mu$sec

($\rho$, $\Gamma_0$, $c_0$, and $sm_0$) listed in Table I, and incorporated a pressure-dependent yield strength, as shown in Fig. 3 and expressed in Eq. 4. Notably, omitting this pressure-dependent yield strength results in significant deviations from the experimental observations, even after parameter calibration. Ruiz-Ripoll, Riedel, Stocchi, et al. (2023)[13] similarly reported that incorporating pressure-dependent yield stress (Drucker–Prager linear strength) and a pressure-dependent shear modulus improves dynamic stress–strain predictions. To capture pressure sensitivity, we therefore adopted the former, i.e., a sigmoidal $P-Y$ (pressure–yield) relation. In our simulations, the volumetric response is decoupled from the deviatoric stress response: the $P-\alpha$ formulation modifies the pressure response through the equation of state to capture porosity evolution, while the $P-Y$ curve governs the shear strength and inelastic strain. This continuous sigmoidal form ensures smooth yield variation under shock loading, consistent with continuum assumptions.

$$Y(P) = a + \frac{b}{1 + e^{-\frac{P-c}{d}}} \tag{4}$$

The parameters of Eq. 4 were calibrated to the 295 m/s weak-shock experiment, with the baseline yield strength set to $a = 0$. The transition parameter $c$ was assigned a value of $1.9 \times 10^{-3}$ Mbar, chosen based on the expected yield-pressure range of sugar grains, providing a reasonable transition scale for pore collapse, rather than representing a directly measured granular yield pressure. The remaining parameters, $b = 0.132$ and $d = 1 \times 10^{-4}$ Mbar, were refined within physically reasonable bounds through parametric adjustment to match the measured particle-velocity history. It should be noted that a non-zero baseline yield strength parameter $a$ in Eq. 4 is generally more representative of granular materials, as such materials typically exhibit a finite yield stress even at vanishing confining pressure. While we acknowledge that $a$ should, in principle, be non-zero, the present study sets $a$=0 because the estimated yield strength at zero confining pressure is negligible compared to the pressure-dependent yield strengths considered in our simulations, thereby maintaining consistency between the model parameters and the simulated conditions. The best match was achieved using this $P$ -$Y$ curve, with a shear modulus of 0.1 Mbar and crush-out pressure (pressure scale) of $1.18 \times 10^{-3}$ Mbar. These values fall within a physically reasonable ranges listed in Table I. This pressure scale is chosen to be of the same order as the grain yield stress (1.1 kbar[6]).

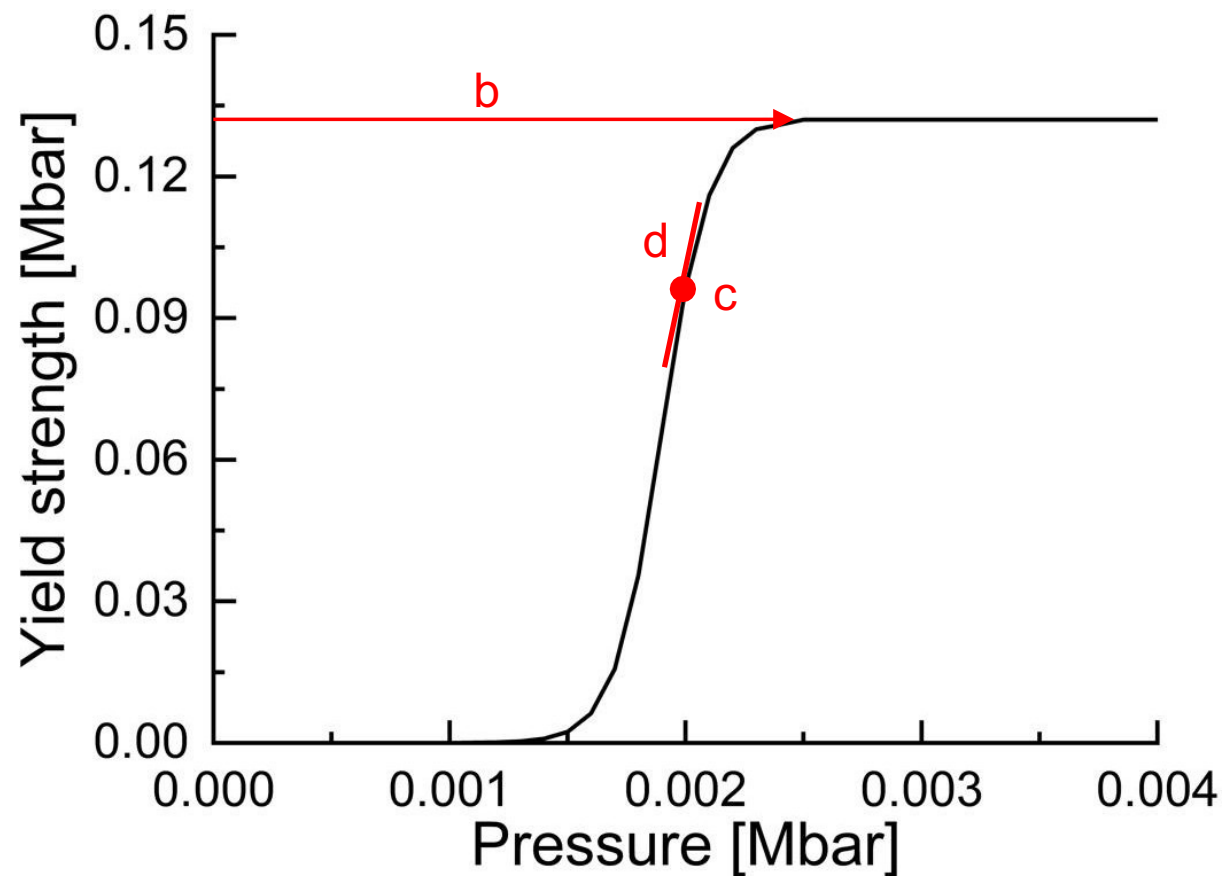

FIG. 3. Pressure-dependent yield strength

The pressure-dependent yield of the strength relation allows the $P$ -$\alpha$ Menko model to effectively capture granular material behavior. Under dynamic loading, particle rearrangement dominates at low pressure, transitioning to particle interlocking at higher pressures. As shown in Fig. 4a, the transmitted wave velocity (in red) gradually increases, reflecting the rise in the yield strength. However, an initial peak appears (in green-colored box) in the transmitted wave unlike in experiments, which is likely induced by the $P$ -$\alpha$ model's tendency to collapse porosity too rapidly in the inelastic regime due to the lack of microinertia, leading to over-prediction of the pressure. A strength model incorporating a micro-inertia term or a time-dependent porosity evolution is likely to regularize the porosity collapse mitigating this initial peak artifact in the velocity, potentially improving model accuracy.

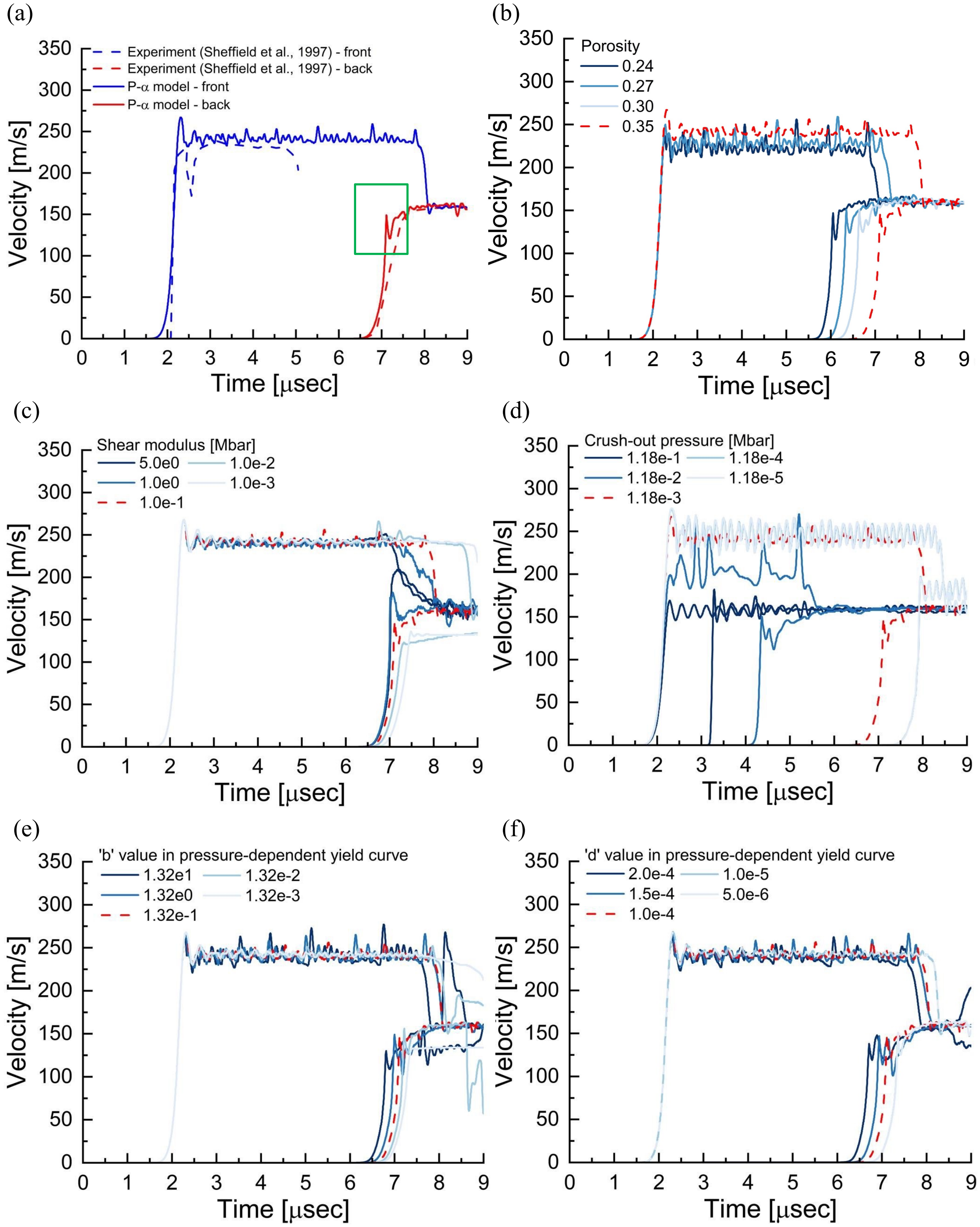

FIG. 4. (a) Comparison of shock velocities between $P$-$\alpha$ model and experimental data (b) $P$-$\alpha$ model with porosity of 0.24, 0.27, 0.30, and 0.35. Sensitivity analysis of $P$-$\alpha$ Menko model with respect to (c) shear modulus, (d) crush-out pressure, $P_c$, (e) yield strength, and (f) slope of the $Y(P)$

Porosity plays a critical role in governing the onset of transmitted velocity, as illustrated in Fig. 4b. While various material parameters will be shown to influence the magnitude of particle velocities (Fig. 4), in this calibrated weak-shock case, porosity predominantly affects the initiation phase of the transmitted velocity evolution, rather than the stiffness or the asymptotic value associated with the plateau state. On the other hand, the slope of the velocity profile is primarily dictated by material parameters, discussed in the following.

A parametric study was conducted to evaluate how model parameters affect particle velocity evolution, and results are compared to the experimental data as the baseline truth. Figure 4 shows the influence of the shear modulus, crush-out pressure ($P_c$), and pressure-yield curve parameters $b$ (maximum yield stress) and $d$ (yield-pressure slope), noting that the broad $P_c$ range is intended for sensitivity analysis rather than representing realistic material values.

The red dashed line marks the parameter set used for the curve deemed suitable to simulate the experimental data in Fig. 4a. Shear modulus influences the onset and magnitude of the transmitted wave velocity; lower values delays shock transmission and reduces peak velocity, likely due to higher deformation and energy dissipation (Fig. 4c)[14]. In Fig. 4d, lower values of $P_c$ delay the transmitted-wave onset, indicating that more energy is dissipated early during pore collapse before a sustained stress wave can form. However, once compaction is achieved, the particle-velocity plateau becomes higher for lower $P_c$, reflecting the fact that the material reaches a denser, stiffer state earlier in the loading path. This behavior highlights that $P_c$ governs the balance between early-time dissipation and the subsequent development of a stiffer, more solid-like response. Parameters $b$ and $d$ also influence shock propagation by modifying plasticity and energy dissipation (Figs. 4e and f). Overall, solid-like behavior parameters are key to accurately modeling granular shock response in the $P$ -$\alpha$ framework. The parameters used in the $P$ -$\alpha$ model are summarized in the Supplementary, along with their methods of determination.

The mesoscale model employs an elasto–perfectly plastic constitutive response for individual grains, with the best fit shown in Fig. 5a. This formulation is appropriate for crushed brittle solids under confined, high-pressure dynamic loading, where hardening effects are secondary and the dominant behavior is elastic deformation followed by yielding[4]. A constant grain yield strength of $Y_0 = 1.1 \times 10^{-3}$ Mbar was adopted based on the reference value in Table I, with the mesoscale model parameters summarized in the Supplementary. Overall, the results show that intrinsic material parameters have a limited influence on the macroscopic evolution of wave velocities. As illustrated in Fig. 5c, variations in the grain-level shear modulus have little effect on the transmitted wave velocity. This behavior contrasts with the continuum $P-\alpha$ Menikoff model (Fig. 4c), where the shear modulus strongly influences the weak-shock response. The difference reflects the distinct physics captured by each approach: in continuum models, the shear modulus acts as an effective stiffness that aggregates unresolved force-chain and grain-interaction effects, directly affecting wave transmission. In contrast, mesoscale simulations explicitly resolve force-chain formation, particle rearrangement, and porosity evolution, so the macroscopic wave speed is governed primarily by geometric packing and compaction rather than by the intrinsic grain modulus. Grain yield strength affects the initiation time of the transmitted wave, but the response is not monotonic, oscillating between 5.5 and 6.5 $\mu$s (Fig. 5d). Aside from this timing variation, yield strength has minimal impact on the overall wave profile. Porosity is the only parameter that continuously modulates the transmitted-wave onset (Fig. 5b), highlighting its dominant influence on the macroscopic response[15]. This trend is consistent with prior mesoscale studies showing that shock behavior is governed primarily by microstructural features[16], and that macroscopic wave speeds are often only weakly sensitive to most intrinsic material parameters[17].

The mesoscale model captures the overall macroscopic response of the transmitted wave in good agreement with the experimental data. Although the slope of the increasing velocity is not matched as closely as in the $P-\alpha$ Menko model, the mesoscale simulations capture characteristic fluctuations in the pre-shock regime (before 6$\mu$s) of the transmitted wave velocity (Fig. 1b and Fig. 5a). These fluctuations, which likely originate from heterogeneous contact interactions and local porosity variations, are also evident in the experimental traces but are not represented in the $P-\alpha$ model.

As shown in Figs. 4 and 5, the numerical simulations capture the overall evolution of the particle velocity but exhibit a late decline compared to experimental traces (Fig. 1b). In the coarse sugar data reported by Sheffield, Gustavsen, and Alcon. (1998)[5], a sharp early velocity drop appears in the particle-velocity trace, yet no corresponding feature is present in the input gauge signal and the original authors did not comment on its origin. We therefore

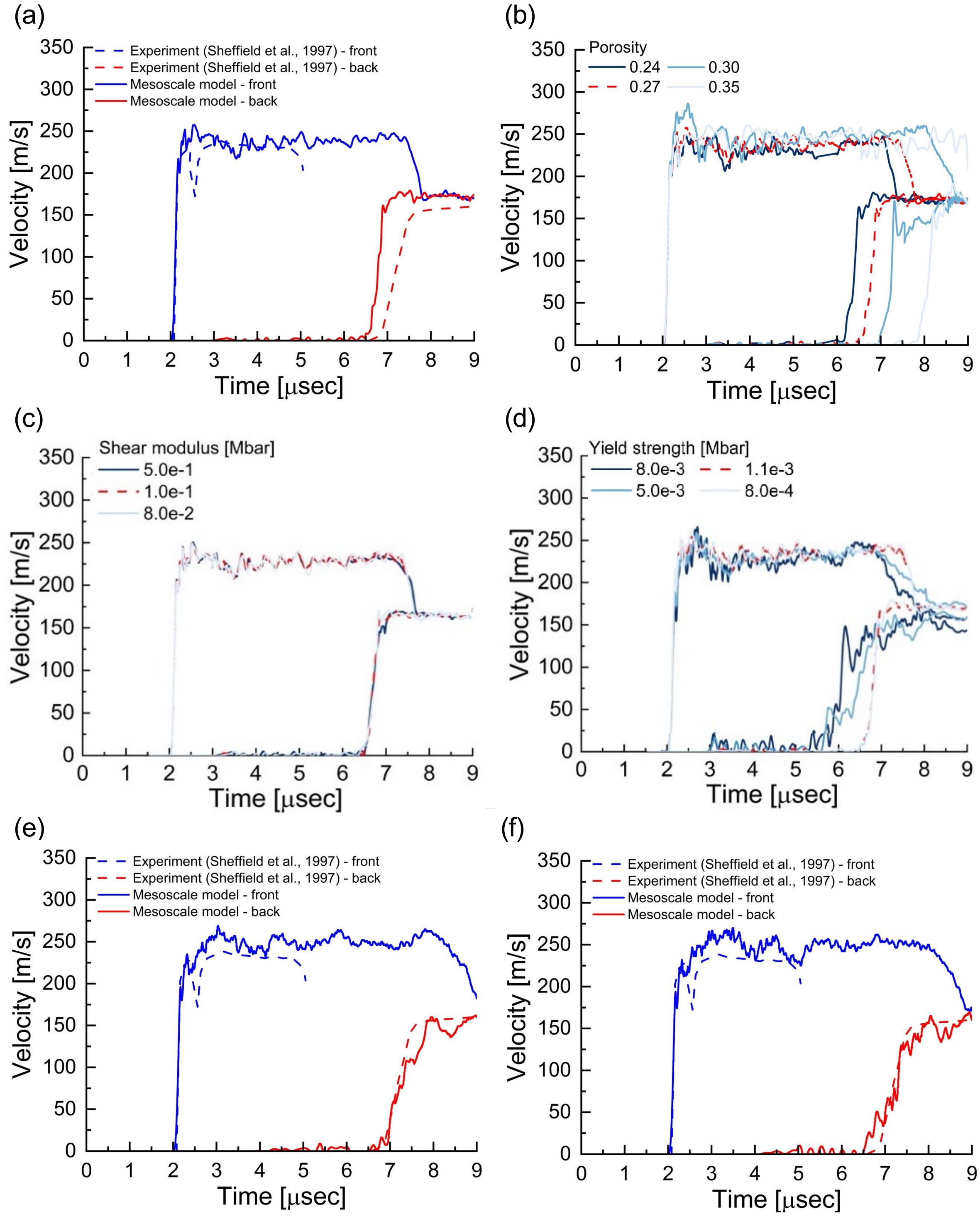

FIG. 5. (a) The best fitting of mesoscale model with elasto-perfectly plastic model to experimental data, and (b) mesoscale model with porosity of 0.24, 0.27, 0.30, and 0.35. Sensitivity analysis of mesoscale model (c) shear modulus, and (d) yield strength. Mesoscale model with elasto-plastic behavior and linear hardening for material response: (e) Fit 1, and (f) Fit 2

interpret this early decrease as an experimental artifact, likely related to gauge–material coupling or signal degradation, and do not attempt to reproduce it in our modeling. In contrast, the fine-sugar data exhibit a later velocity reduction ($\approx 7\mu sec$) that is consistent with our simulations. This interpretation supports the consistency of our approach while recognizing possible experimental anomalies in the coarse sugar data.

We additionally examined the consequence of linear hardening on mesoscale grain response. Figs. 5e and 5f show best-fit results using an elasto-plastic model with linear hardening, emphasizing the alignment of overall trends with experiments, particularly for the transmitted wave. Assuming that crushable solids exhibit stress hardening at the continuum scale, isotropic elastic-plastic hardening may approximate grain behavior under confined compression[12]. The linear hardening parameter helps modulate weak shock waves and offers an effective way to tune shock speeds in 2D simulations to match 3D experimental data, precluding any requirement for porosity conversion Eq. 3. While linear hardening effectively captures particle velocity trends well, multiple parameter sets yield similarly good fits. For example, Fig. 5e uses a shear modulus of 0.1 Mbar, flow strength of 0.004 Mbar, and yield strength of 0.005 Mbar, while Fig. 5f corresponds to 0.1 Mbar, 0.005 Mbar, and 0.01 Mbar, respectively. Additional experimental data may facilitate constraining the parameters.

This study examines a single weak-shock condition (295 m/s) to calibrate and compare the 2D continuum-based $P-\alpha$ Menikoff model with mesoscale simulations, using particle-velocity waveforms from gas-gun experiments by Sheffield, Gustavsen, and Alcon.[5]. The $P-\alpha$ model required a pressure-dependent sigmoidal yield function to match the experimental response but exhibited an early peak in transmitted velocity due to the absence of a microinertia term for pore collapse. Its predictions were highly sensitive to the crush-out pressure, which governs the transition from pore-collapse to solid-like behavior. In contrast, the mesoscale simulations matched the transmitted wave by adjusting porosity to account for 2D-3D differences and showed reduced sensitivity to parameter variations. Porosity emerged as the dominant factor controlling wave onset, underscoring its primary role in the macroscopic response of granular media. Although both modeling approaches matched the experimental data, they rely on distinct physical interpretations. The continuum $P-\alpha$ model employs effective parameters, such as shear modulus, crush-out pressure, and pressure-dependent yield strength, to represent unresolved grain-scale physics, whereas the mesoscale model directly resolves particle packing and pore evolution (further details are provided in the Supplementary). By explicitly capturing force-chain stiffness and compaction energetics, the mesoscale simulations provide a physical basis for interpreting the continuum shear modulus and crush-out pressure $P_c$ as coarse-grained representations of grain-scale mechanics. While the present results are restricted to this calibrated loading state, extending the framework to multiple driver velocities and leveraging mesoscale modeling to elucidate the physical origins of continuum-model parameters represent important directions for future work.

**Data Availability Statement:** CSV files containing the experimental particle-velocity histories and the corresponding continuum and mesoscale simulation data are available from the lead author upon reasonable request.

## SUPPLEMENTARY MATERIALS

The supplementary material includes additional descriptions of (A) the experimental setup from Sheffield, Gustavsen, and Alcon. (1998)[5], (B) the implementation of $P - \alpha$ modeling in continuum models, (C) mesoscale modeling in FLAG, and (D) cross-scale interpretation, which support the descriptions of the numerical methods and model parameters used in this study.

## ACKNOWLEDGMENTS

Research on geomechanics aspects and mesoscale analysis under shock compaction was supported by the Laboratory Directed Research and Development program of Los Alamos National Laboratory (LANL) under project number 20220811PRD4 awarded as part of the Laboratory's Director's Fellowship. Research on computational modeling using FLAG and validation of simulations was supported by U.S. Department of Energy's ASC/PEM/Enabling Manufacturing through LANL. LANL is operated by Triad National Security, LLC, for the National Nuclear Security Administration of U.S. Department of Energy (Contract No. 89233218CNA000001). Authors appreciate discussions with Drs. C. Rousculp, C. Ticknor, A. Koskelo and C. Scovel from LANL.

# A Comparison between Separately Calibrated $P$-$\alpha$ and Mesoscale Models for Weak Shock Compaction of Granular Sugar

## Supplementary Material

### A. A gas gun experimental setup of Sheffield et al. (1998)[1]

This study utilizes the experimental setup of Sheffield et al.[1], illustrated schematically in Fig. 1a, where experiments were performed using a gas gun to launch projectiles composed of polychlorotrifluoroethylene (Kel-F), which struck a Kel-F containment cell (outer diameter: 68.6 mm; inner diameter: 40.6 mm) filled with sugar powder. The simulated granular sugar sample in Fig. 2a was pressed into a compact and confined between the Kel-F impact face and a cylindrical polymethylmethacrylate (PMMA) backing plug. The compacted sugar layer was 4 mm thick and had a density of 65% of its theoretical maximum (TMD). The tests were carried out at an impact velocity of 295 $m/s$. Sheffield et al. mounted magnetic particle velocity gauges on the Kel-F and PMMA plastic cell in contact with the granular sugar sample. Magnetic velocity gauges were used to measure the particle velocities traveling from the impactor to backing plug through granular sugar shown in Fig. 1a, determining the shock input and wave dispersion response.

### B. Implementation of $P$-$\alpha$ model into continuum models

In continuum shock-physics models using the P-$\alpha$ model, the material description is split into two essential parts: the solid-matrix equation of state (EOS) and porous-material properties. The thermodynamic response of the material is provided entirely by a solid-matrix equation of state (EOS), which defines how the solid material (and equivalently, fully dense) responds to compression, heating, and shock loading. The $P$-$\alpha$ model is layered on top of this EOS to describe the influence of porosity by specifying the initial distension, the pressures at which pores begin evolving and accomplish complete collapse, and the shape of the compaction curve. During each timestep, the solver evaluates the matrix pressure from the dense-material EOS and then scales it using the current distension to obtain the effective porous pressure. As the shock compresses the material, the $P$-$\alpha$ model updates distension, $\alpha$ (Eq 1), to represent compaction, and this evolution modifies density, volumetric stiffness, and affects how compaction work is partitioned into internal energy depending on the chosen closure. In classical P–$\alpha$ formulations, pore collapse is commonly treated as irreversible in the sense that distension decreases during loading and does not recover on unloading. However, the associated energy bookkeeping is implementation dependent: the mechanical work done during crush may be represented as dissipated heating, as an additional compaction-energy (potential-like) contribution to the internal energy, or implicitly through plastic work in the accompanying strength model. Consequently, the P–$\alpha$ model does not, by itself, uniquely prescribe a single "irreversible crush-work" term added to the specific internal energy; rather, the conversion of compaction work to internal energy depends on the chosen closure and constitutive models. By contrast, the Menikoff–Kober reformulation embeds compaction within a thermodynamically consistent EOS framework by augmenting the solid free energy with a compaction potential and determining the equilibrium solid volume fraction from a free-energy condition; in this view, the energetic effect of compaction is accounted for through the EOS (via the compaction potential) rather than being introduced as an ad hoc dissipative crush-work term.. Thus, the solid-matrix EOS provides the fundamental shock physics, while the $P$-$\alpha$ model adjusts that response to account for pore collapse, allowing the continuum model to represent the transition from a highly compressible porous solid to a fully dense material under shock loading.

In the $P$-$\alpha$ simulation using the Menikoff-Kober[2] model, particle velocity was recorded at the front and rear face centers of the pressed granular sugar specimen using a modeled VISAR (Velocity Interferometer System for Any Reflector) as shown in Fig. 1a. In FLAG, VISAR was simulated by tracking velocity at specific nodes at the same positions as those of the reflectors in the experiments, and aligned with the experimental setup's line of sight. Here, we define the initial velocity rise at the front face of the sugar specimen as the input particle velocity and the signal at the rear sugar face (sugar/PMMA interface) as the transmitted wave velocity.

#### 1. Classical P–$\alpha$ (Herrmann / Carroll–Holt implementation)

Specifically, the classical $P$-$\alpha$ model[3] calculates a distension parameter $\alpha = V/V_s = \rho_s/\rho$, where $\rho$ is the bulk (porous) density and $\rho_s$ is the solid density. The model first computes the matrix pressure $P_s(\rho_s, E)$ from a fully dense EOS, and then scales it to obtain the porous pressure $P = P_s/\alpha$, per Carroll & Holt[4]. During loading, the solver updates $\alpha$ using a prescribed crush curve (one common choice is the piecewise-linear form in Eq. (S1)) defined by the initial distension $\alpha_0$, an elastic compaction threshold $P_e$, and a saturation pressure $P_{sat}$.

$$\alpha(P) = \begin{cases} \alpha_0, & P \le P_e, \\ \alpha_0 - (\alpha_0 - 1)\dfrac{P - P_e}{P_{\mathrm{sat}} - P_e}, & P_e < P < P_{\mathrm{sat}}, \\ 1, & P \ge P_{\mathrm{sat}}. \end{cases} \tag{S1}$$

Porosity remains unchanged for $P < P_e$, collapses progressively for $P_e < P < P_{\mathrm{sat}}$, and vanishes when $P \ge P_{\mathrm{sat}}$. This updated $\alpha$ is then used to adjust density and specific volume consistently, ensuring that compaction is irreversible (i.e., $\alpha$ is constrained to be non-increasing on unloading) and that the material approaches the fully dense state as pores close. Compaction work is converted into internal energy through the solver's standard energy update (and any accompanying strength/closure model), thereby influencing the thermal response; the classical P–$\alpha$[3,4] formulation specifies the porous EOS and the distension evolution, while the detailed partition of compaction work into heating is implementation dependent. The resulting porous pressure and density feed into the momentum and energy conservation equations, while deviatoric stresses are handled by a separate strength model. In this way, the $P - \alpha$ formulation allows continuum models to capture the transition from highly compliant, pore-dominated response to the stiffer behavior of the compacted solid under shock loading.

### 2. Menikoff–Kober (MK) thermodynamic $P$–$\alpha$ formulation

The MK formulation is a thermodynamically consistent framework for porous materials whose weak-shock Hugoniot locus is concave in the ($u_p$, $u_s$)-plane[2]. Rather than prescribing distension collapse through a piecewise crush curve, MK constructs the porous response from a Helmholtz free energy of the form $\Psi(V, T, \chi) = \Psi_s(V_s, T) + B(\chi)$, where $V_s = \chi V$ is the solid specific volume and $B(\chi)$ is a compaction potential[2]. In the MK formulation, $\chi$ denotes the solid volume fraction (so $\alpha = 1/\chi$). The porous (macroscopic) pressure is obtained from the solid EOS and scaled by the solid fraction, $P = \chi P_s$ (equivalently $P = P_s/\alpha$). Unlike the classical P–$\alpha$ approach where $\alpha$ is prescribed directly as a crush curve $\alpha(P)$, MK determines $\chi$ from a thermodynamic equilibrium condition, which yields an implicit equilibrium relation $\chi = \chi_{eq}(V_s P_s)$ rather than a purely pressure-based crush curve[2]. A simple and useful empirical choice for $\chi_{eq}$ introduces a pressure-scale parameter $P_c$ that controls the rate/curvature of compaction[2].

In practice, given the specific internal energy E and density, ($\rho$, $E$), at a time step, the MK update is implemented by solving implicitly for the equilibrium solid volume fraction $\chi$ consistent with the MK equilibrium condition and the underlying solid EOS; the resulting solid pressure $P_s$ and porous pressure $P = \chi P_s$ are then returned to the hydro solver. The EOS includes an explicit compaction contribution through $B(\chi)$, and the energetic effects of pore collapse are accounted for through the EOS structure rather than through an ad hoc "crush-work" term[2].

TABLE I. P-$\alpha$ model parameters

| Constant | Value | Units | Reference/rationale/calibration |
| --- | --- | --- | --- |
| $\rho_s$ | 1.58 | g/cm$^3$ | Single particle density (Trott et al., 2007[5]; Sheffield et al., 1998[1]) |
| $\Gamma_0$ | 1.04 | - | Grüneisen parameter (Trott et al., 2007[5]) |
| $c_0$ | 0.304 | cm/$\mu$s | Sound speed (Trott et al., 2007[5]) |
| $\phi$ | 0.35 | - | Porosity (Sheffield et al., 1998[1]) |
| $sm_0$ | $1.0 \times 10^{-1}$ | Mbar | Shear modulus; estimated using bulk modulus and void ratio (Trott et al., 2007[5]) |
| $a$ | 0 | Mbar | Baseline yield strength; zero-pressure yield strength assumed negligible |
| $b$ | 0.132 | Mbar | Maximum pressure-dependent yield strength $Y(P)$; calibrated |
| $c$ | $1.9 \times 10^{-6}$ | Mbar | Transition parameter; calibrated |
| $d$ | $1.0 \times 10^{-4}$ | Mbar | Slope of pressure-dependent yield strength $Y(P)$; calibrated |
| $P_{s,\mathrm{ref}}$ | $1.0 \times 10^{-6}$ | Mbar | Reference-state solid pressure (atmospheric pressure) |
| $v_{s,\mathrm{ref}}$ | 0.633 | cm$^3$/g | Reference-state solid specific volume (initial solid volume in our model) |
| $\chi_{\mathrm{ref}}$ | 0.65 | - | Reference-state solid volume fraction (= $1/\alpha$) (Sheffield et al., 1998[1]) |
| $P_c$ | $1.18 \times 10^{-3}$ | Mbar | Crush-out pressure; calibrated |

Following MK, porosity effects are most important at pressures below the pure-solid yield strength, a regime often referred to as the crush-up pressure[2]. In this work we refer to the fitted MK pressure-scale parameter as a *crush-out pressure* $P_c$ to emphasize that it sets the characteristic compaction scale; however, $P_c$ should be interpreted as controlling the smooth approach to the fully dense limit rather than as a sharp cutoff for complete pore collapse.

Table I summarizes the parameters used in the P-$\alpha$ model and describes how each value was determined. Parameters were taken from the literature, derived from experimental data, estimated from literature-based values with minor fitting and condition-dependent conversions, or determined purely through calibration to experimental data. Where necessary, reported ranges were used as guidance and adjusted slightly to improve agreement with observations, with final refinements made to match the experimental results.

## C. Mesoscale modeling in FLAG

FLAG is a multiphysics hydrocode developed at Los Alamos National Laboratory (LANL) for modeling shock dynamics simulations in solids and fluids. It solves the continuum momentum and energy equations using a finite volume formulation and, in the Lagrangian method, tracks the motion of a computational mesh that moves with the material. In this approach, each grain is represented by a deformable Lagrangian region whose boundaries govern mechanical interaction with other grains, allowing force transmission while preventing interpenetration.

In this work, we apply and extend this mesoscale modeling framework to granular sugar. Individual grains are represented as deformable Lagrangian regions described by an isotropic elastic–plastic constitutive law with linear hardening and a Mie–Grüneisen equation of state. FLAG computes grain–grain and grain–boundary interactions using surface-to-surface contact algorithms, and in the present simulations these interfaces are treated as frictionless. Furthermore, neglecting friction isolates the role of pore collapse and porosity evolution in weak-shock compaction, and is consistent with previous stud-ies. They have demonstrated that compaction can be effectively modeled without explicitly incorporating fracture or frictional contact[6,7]. Borg and Vogler (2009)[8] showed that, under low levels of deformation, energy dissipation associated with viscoplastic work dominates over dissipation from grain–grain contact and friction. Consistent with this observation, the present simulations do not employ a fracture model. Although a continuum-based "$P_{min}$" failure model, similar to that used by Borg and Vogler (2009)[8], could be introduced, such an approximate treatment can introduce uncertainty in the post-failure response. In particular, abruptly forcing the shear and deviatoric stresses to zero may violate energy conservation and reduce the physical fidelity of the simulation. The approach adopted in this study therefore follows the FLAG mesoscale methodology validated for granular salt[9].

In this study, square-shaped grains (0.120 mm equivalent diameter) were stacked through dry pluviation, modeled to generate gravity-driven particle packings analogous to the experimental procedure in which grains are rained into a hollow cylinder and then lightly compressed to reach a target tap density. The simulations were performed using an integrated workflow combining MATLAB and LAMMPS, as described in Seo et al. (2024)[9]. Because the pluviation process typically produced packings with tap densities higher than desired, a selective particle-removal strategy was applied to adjust the porosity while preserving particle contacts. Similar deletion or insertion approaches have been used previously to tailor granular configurations[10]. In the present work, particles chosen for removal were those with relatively high coordination numbers, following the method detailed in Seo et al. (2024)[11].

This study employs 2D mesoscale simulations, a common approach used to avoid the substantial computational cost associated with resolving full 3D grain interactions and porosity in deformable granular systems[8]. Although mapping 3D porosity to an equivalent 2D value does not recover the full 3D contact network, 2D mesoscale simulations remain widely used because of their substantially lower computational cost. Borg and Vogler (2013)[12] showed that 2D configurations can reproduce key trends in stress, particle velocity, and temperature, though they generally exhibit lower coordination numbers and softer response than their 3D counterparts. Consistent with these observations, we employ a stereological mapping based on relative density (Eq. 3) to construct a 2D packing with porosity equivalent to the 3D system. This approach preserves the volumetric compaction behavior while acknowledging the reduced force-chain connectivity inherent to 2D assemblies, and has been validated in prior FLAG[9] for predicting shock-compaction trends.

Table II summarizes the parameters used in the mesoscale model.

TABLE II. Mesoscale model parameters

| Constant | Value | Units | Reference/rationale/calibration |
|---|---|---|---|
| $\rho_s$ | 1.58 | g/cm$^3$ | Single particle density (Trott et al., 2007[5]; Sheffield et al., 1998[1]) |
| $\Gamma_0$ | 1.04 | - | Gru¨neisen parameter (Trott et al., 2007[5]) |
| $c_0$ | 0.304 | cm/$\mu$s | Sound speed (Trott et al., 2007[5]) |
| $\phi$ | 0.27 | - | Porosity; rationale (converted from 3D to 2D based on Seo et al., 2024a[9]) |
| $sm_0$ | $1.0 \times 10^{-1}$ | Mbar | Shear modulus; estimated using bulk modulus and void ratio (Trott et al., 2007[5]) |
| $\sigma_y$ | $1.1 \times 10^{-3}$ | Mbar | Yield stress (Trott et al., 2007[5]) |

## D. Cross-Scale Interpretation

To establish a direct physical bridge between the mesoscale simulations and the continuum $P-\alpha$ formulation, we introduce a detailed cross-scale interpretation grounded in the mechanics resolved at the grain level. In the mesoscale model, the stiffness of the granular assembly arises from the formation and evolution of force chains, the coordination number of the contact network, and the progressive collapse of pore space under loading. These mechanisms provide a physical basis for interpreting the continuum shear modulus as an effective stiffness that reflects the averaged response of the force-chain network rather than a true intrinsic material constant of the solid grains. Similarly, the continuum 'crush-out pressure' $P_c$ can be interpreted as a characteristic compaction pressure scale, i.e. the pressure range over which mesoscale simulations show the most rapid reduction in pore volume (and associated increase in solid contact force chains), as evidenced by the spatial progression of pore collapse and the partitioning of input energy into compaction versus elastic/plastic deformation. The mesoscale results also capture stress and velocity fluctuations arising from heterogeneous particle arrangements, features that are inherently homogenized in the continuum representation and absorbed into its effective parameters. By relating mesoscale force-chain stiffness, compaction energetics, and heterogeneity-driven fluctuations to their continuum counterparts, this interpretation provides a mechanistic foundation for the effective parameters used in the $P-\alpha$ model and demonstrates how grain-scale physics manifests at the continuum scale.